\documentclass{aa}
\usepackage{graphicx,epsfig}
\usepackage{natbib}
\usepackage[usenames]{color}
\usepackage{txfonts}
\usepackage{url}
\usepackage[dvips]{hyperref}
\usepackage{stfloats}
\usepackage{threeparttable}

\newcommand{\HeII}{\ion{He}{ii}}

\newcommand{\FeX}{\ion{Fe}{x}}
\newcommand{\FeIX}{\ion{Fe}{ix}}

\newcommand{\kms}{km~s$^{-1}$}
\newcommand{\degree}{\ensuremath{^\circ}}

\begin{document}

\title{STEREO quadrature observations of coronal dimming at the onset of mini-CMEs}

\author{D.~E. Innes\inst{1} \and S.~W. McIntosh\inst{2} \and A. Pietarila\inst{1}}
\institute{Max-Planck Institut f\"{u}r Sonnensystemforschung, 37191 Katlenburg-Lindau, Germany \and High Altitude Observatory, National Center for Atmospheric Research, P.O. Box 3000, Boulder, CO 80307, USA}
\offprints{D.E. Innes \email{innes@mps.mpg.de}}

\date{Received ...; accepted ...}

\abstract
{Using unique quadrature observations with the two STEREO spacecraft, we investigate coronal dimmings at the onset of small-scale eruptions. In CMEs they are believed to indicate the opening up of the coronal magnetic fields at the start of the eruption. }
{It is to determine whether coronal dimming seen in small-scale eruptions starts before
or after chromospheric plasma ejection.}
 {One STEREO spacecraft obtained high cadence, 75~s, images in the \HeII\ 304\AA\ channel, and the other
 simultaneous images in the \FeIX/\FeX\ 171\AA\ channel. We concentrate on two well-positioned chromospheric eruptions
 that occurred at disk center in the 171\AA\ images, and on the limb in 304\AA. One was in the quiet Sun and the other was in an
 equatorial coronal hole.
  We compare the timing of chromospheric eruption seen in the 304\AA\ limb images with the
 brightenings and dimmings seen on disk in the 171\AA\ images.
Further we use off-limb images of the low frequency 171\AA\ power to infer the coronal structure near the eruptions. }
{In both the quiet Sun and the coronal hole eruption, on disk 171\AA\ dimming was seen before the chromospheric eruption, and in both cases it extends beyond the site of the chromospheric eruption. The quiet Sun eruption occurred on the outer edge of the enclosing magnetic field of a prominence and may be related to a small disruption of the prominence just before the 171\AA\ dimming. }
{These small-scale chromospheric eruptions started with a dimming in coronal emission just like their larger counterparts. We therefore suggest that a fundamental step in triggering them was the removal of overlying coronal field. }


\titlerunning{Mini-CME onset dynamics}
\authorrunning{D.~E. Innes \and S.~W. McIntosh \and A. Pietarila}

\maketitle

\section{Introduction}

It was recently noted that small-scale eruptions from the quiet Sun appear remarkably similar in 171\AA\ EUV image sequences to CMEs, hence the term mini-CMEs. They show deep dimmings and flare-like brightenings at the onset site and, in about one third of the events, more extended fainter dimming waves propagating out to several tens of Mm \citep{Ren08, Innes09, Podlad10}. Furthermore their frequency extends the power-law distribution of large CMEs down to the 20~Mm scale \citep{Schrijver10}.

One important characteristic seen in several well-observed large CMEs, is a coronal dimming just prior to the filament or flare eruption \citep{Zarro99, Sterling04, Harrison03, Gopal06, Harrison07}. This is believed to be due to the opening of the coronal magnetic field holding down the magnetic flux rope of the CME. As such it holds important clues to the onset mechanism \citep{amari99, antiochos99, low03}.

In the quiet Sun, mini-CMEs occur at the junctions of supergranules where the mixed polarity magnetic fields are twisted in the downdrafts of the supergranular flows \citep{Innes09}. The environment is therefore similar to an active region with significant shear.
If the eruptions in the quiet Sun are scaled down versions of CMEs then the eruption onset should also be related to the re-structuring of the corona.

In this letter, we investigate the onset of two small eruptions: one in the quiet Sun and the other in a coronal hole.
First the observations are described, then the data is presented. In the discussion section, we summarize the velocity and timings of the dimming fronts seen in the 171\AA\ and the chromospheric eruptions seen in the 304\AA\ images. Finally we use low frequency power maps of off-limb 171\AA\ intensity to reveal the coronal structure and use these to back-up our conclusion that coronal re-structuring preceded the chromospheric eruptions.

\begin{figure}
\centering
\includegraphics[width=\linewidth]{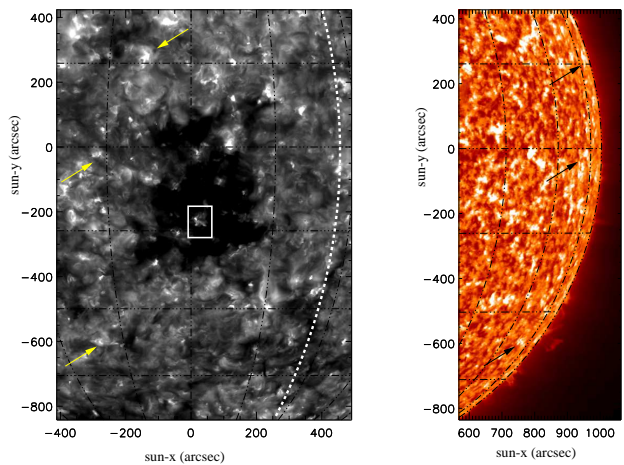}
\caption{The regions observed in the 171\AA\ (left) and 304\AA\ channels (16:30 - 17:00~UT on 14 Feb 2009).
On the 12 Feb the longitude marked with a white dotted line was at disk center. The grid spacing is 15\degree. The arrows point to common structures in the 171\AA\ and 304\AA\ images. The white box surrounds the site of the coronal hole event.} \label{14_av_fig}
\end{figure}

\section{Observations}

The observations were made using the twin EUVI/SECCHI telescopes on the STEREO spacecraft \citep{Howard08}
from 14:00 - 18:00~UT on 12-15 Feb 2009, configured so that one telescope took images in the chromospheric \HeII\ 304\AA\ channel and the other simultaneous images in the \FeIX/\FeX\ 171\AA\ channel. At this time STEREO-B was 47.5\degree\ behind the Earth and STEREO-A was 43.4\degree\ ahead. Thus the separation angle between the two spacecraft was 91\degree. During the four hours of observation on each day, STEREO-B took images approximately every 75~s in the 304\AA\ channel while STEREO-A obtained images in the 171\AA\ channel. Intervening images with the opposite configuration (i.e. STEREO-A 304\AA\ and STEREO-B 171\AA) were taken with a lower cadence, 150~s, and in addition both telescopes obtained context images in
the 195Å and 284Å channels every 10~min.

The data have been calibrated using the standard solarsoft routines\footnote{sohowww.nascom.nasa.gov/solarsoft}. They were corrected for the telescope roll angle and deprojected so that the solar B angle is zero. The disk center images were then corrected for differential rotation, and aligned to the middle time of each sequence. We then expanded the Sun in the B images to match the radius of the A images. After using the numbers from the headers, we still found small corrections were necessary to achieve perfect coalignment of features. The Sun center in the STEREO-B images was shifted north by 5\arcsec.

We concentrate on two events seen in the high cadence data that appeared at the west limb in the 304\AA\ and at disk center in 171\AA. One occurred in the quiet Sun on the 12 Feb. The other was in an equatorial coronal hole on the 14 Feb. Both events are within longitude $\pm15\degree$ of the disk center in STEREO-A. For each event, we have created two animations. One shows the 171\AA\ disk center intensity with the simultaneous 304\AA\ limb image alongside and the other shows the corresponding running difference 171\AA\ next to the limb in 304\AA. The running difference is the logarithm of the intensity ratio at the specified time and six minutes earlier. Thus on a linear scale comparable dimmings and brightenings have similar contrast. The coloured movies and images use a gamma factor of 0.5 and the black and white images have a gamma factor of 1.0.

Images of the region as seen from the two STEREO spacecraft on 14 Feb 2009 (day of the coronal hole eruption) are shown in Fig.~\ref{14_av_fig}. As can be seen, on this day the coronal hole was at disk center in STEREO-A and therefore on the limb in B. The Sun rotates roughly 13\degree\ per day, so two days earlier, when the quiet Sun eruption was observed, the Sun along longitude 26\degree\ was at disk center in 171\AA. Most of the bright patches east of the central meridian in the the 171\AA\ can be identified as bright regions in the 304\AA. A few of the more obvious ones are indicated with arrows.

\subsection{Quiet Sun eruption}
\begin{figure}
\centering
\includegraphics[width=7.5 cm]{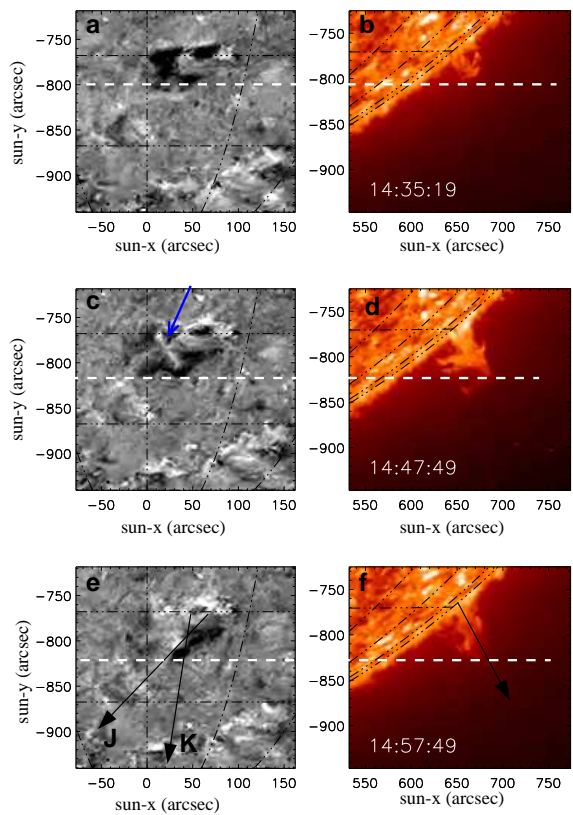}
\caption{STEREO-A 171\AA\ base difference, and simultaneous STEREO-B 304\AA\ intensity images of the quiet Sun eruption: onset (top row), beginning of the fall back (middle row) and later expansion phase (bottom row). {The blue arrow in (c) points to the brightening mentioned in the text}. The grid spacing is 10\degree. The long black arrows in the bottom 171\AA\ and 304\AA\ images indicate the positions and directions of the time series in Fig.~\ref{12_ts_fig}. The evolution is shown in the on-line movies \href{http://www.mps.mpg.de/data/outgoing/innes/dims/vcu_qs.gif}{\tt qs.gif} and \href{http://www.mps.mpg.de/data/outgoing/innes/dims/vcu_qs_r.gif}{\tt qs\_r.gif}} \label{12_cu_fig}
\end{figure}

On the disk in 171\AA\ this eruption looks like a mini-CME (Fig.~\ref{12_cu_fig}). One sees both a dimming and brightening in 171\AA\ at the onset site and a larger scale dimming wave. The 304\AA\ observations on the limb show a surge-like eruption.
 The 171\AA\ images in Fig.~\ref{12_cu_fig} are the intensity difference from the average over the 4 hour observing period (base difference).

\begin{figure}
\centering
\includegraphics[width=7cm]{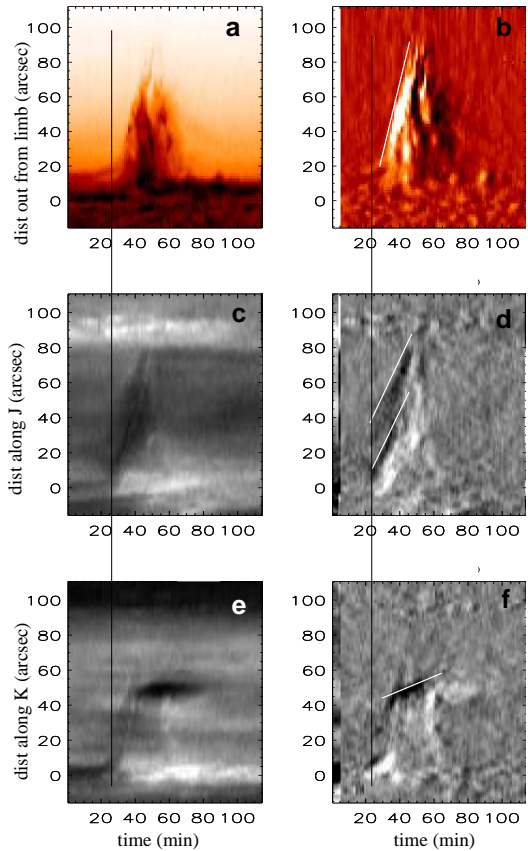}
\caption{304\AA\ and 171\AA\ time series: STEREO-B 304\AA\ along arrow in {Fig.~\ref{12_cu_fig}}f (a) intensity (negative) (b) running difference, and simultaneous STEREO-A 171\AA\ along arrow { J} in Fig.\ref{12_cu_fig}e (c) base difference (d) running difference, and along arrow K in Fig.~\ref{12_cu_fig}e (e) base difference (f) running difference. The white lines indicate fronts listed in Table 1.} \label{12_ts_fig}
\end{figure}

The frames in Fig.~\ref{12_cu_fig} represent the onset (a and b), the time when the surge starts to fall back (c and d), and the late phase structure (e and f). In all frames, a white dashed line has been drawn at the height of the 304\AA\ off-limb so that it is easy to compare its height with the dimming extension. Time series along the arrows in the bottom frames are shown in Fig.~\ref{12_ts_fig}.
The direction along arrow J cuts through the initial dimming, a subsequent brightening seen in Fig.~\ref{12_cu_fig}c, and the later opening of the dimming region. Arrow K cuts through the erupting chromospheric loop.
The velocities and onset times of the different fronts, marked with white lines in Fig.~\ref{12_ts_fig}, are given in Table 1.

In the top frame of Fig.~\ref{12_cu_fig}, the dimming is {south of the white dashed line}, so these outer parts cannot be dimmed by absorption due to chromospheric material along the line-of-sight. The
 running difference movie,{\tt qs\_r.gif}, 
 shows very clearly that the 171\AA\ dimming is seen before and south of the chromospheric eruption. Here a dark, with bright behind, loop moves south at exactly the same time as the erupting loop is seen above the limb. This is reflected in Figs.~\ref{12_ts_fig}b and f, which show the same slow loop rise in 304\AA\ and 171\AA\ before the main eruption. Further confirmation of the early coronal restructuring is given in Fig.~\ref{12_ts_fig}d which shows two dimming fronts that start simultaneously at two different positions, and almost 10~min before the chromospheric eruption.

When the chromospheric material starts to fall back to the Sun, there is a sudden brightening in 171\AA. The structure is {indicated by the blue arrow} in Fig.~\ref{12_cu_fig}c. The brightening increase seen in the time series plots (Fig.~\ref{12_ts_fig}c-f) between 50 and 60~min is subtle but real. It caused the opening up of the dimming region along arrow {J}.

\begin{figure}
\centering
\includegraphics[width=4cm]{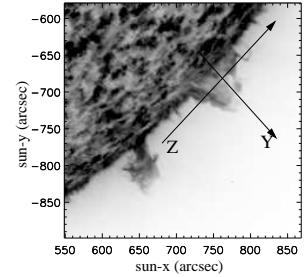}
\caption{STEREO-B 304\AA\ negative intensity image of quiet Sun eruption and neighbouring prominence at 14:45~UT 12 Feb 2009. Arrows Y and Z indicate the position and direction of the times series in Fig.~\ref{12_ts_pr_fig}} \label{cu_12_pr_fig}
\end{figure}

\begin{figure}
\centering
\includegraphics[width=6cm, height=3cm]{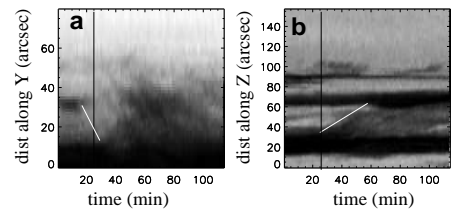}
\caption{Time series of 304\AA\ intensity (negative) along the arrows in Fig.~\ref{cu_12_pr_fig} (a) Y and (b) Z. The black vertical lines are drawn at the same time as the ones in Fig.~\ref{12_ts_fig}. White lines show fronts listed in Table 1.} \label{12_ts_pr_fig}
\end{figure}

As shown in Fig.~\ref{cu_12_pr_fig}, there was a small prominence about 100\arcsec\ northward along the limb from the eruption. The time series along the arrows Y and Z are shown in Fig.~\ref{12_ts_pr_fig}. The time series along Y, pointing out from the limb, shows that part of the prominence disrupts about 5 min before the eruption onset. We have measured a downward velocity of 12~\kms\ (Table~1). There is also a very faint outflow but it is {not} possible to measure its velocity. The most obvious front is seen moving towards the position of the disrupted structure along Z, at the time of the dimming onset in the neighbouring eruption. The 171\AA\ disk images in the movie,,
show a couple of parallel dark channels at the right latitude to be {the} filaments corresponding to the two edges of the prominence. There is a small 171\AA\ brightening between the channels at around 14:17 UT that may be the disk signature of the partial eruption.

\subsection{Coronal hole}
\begin{figure}
\centering
\includegraphics[width=8cm]{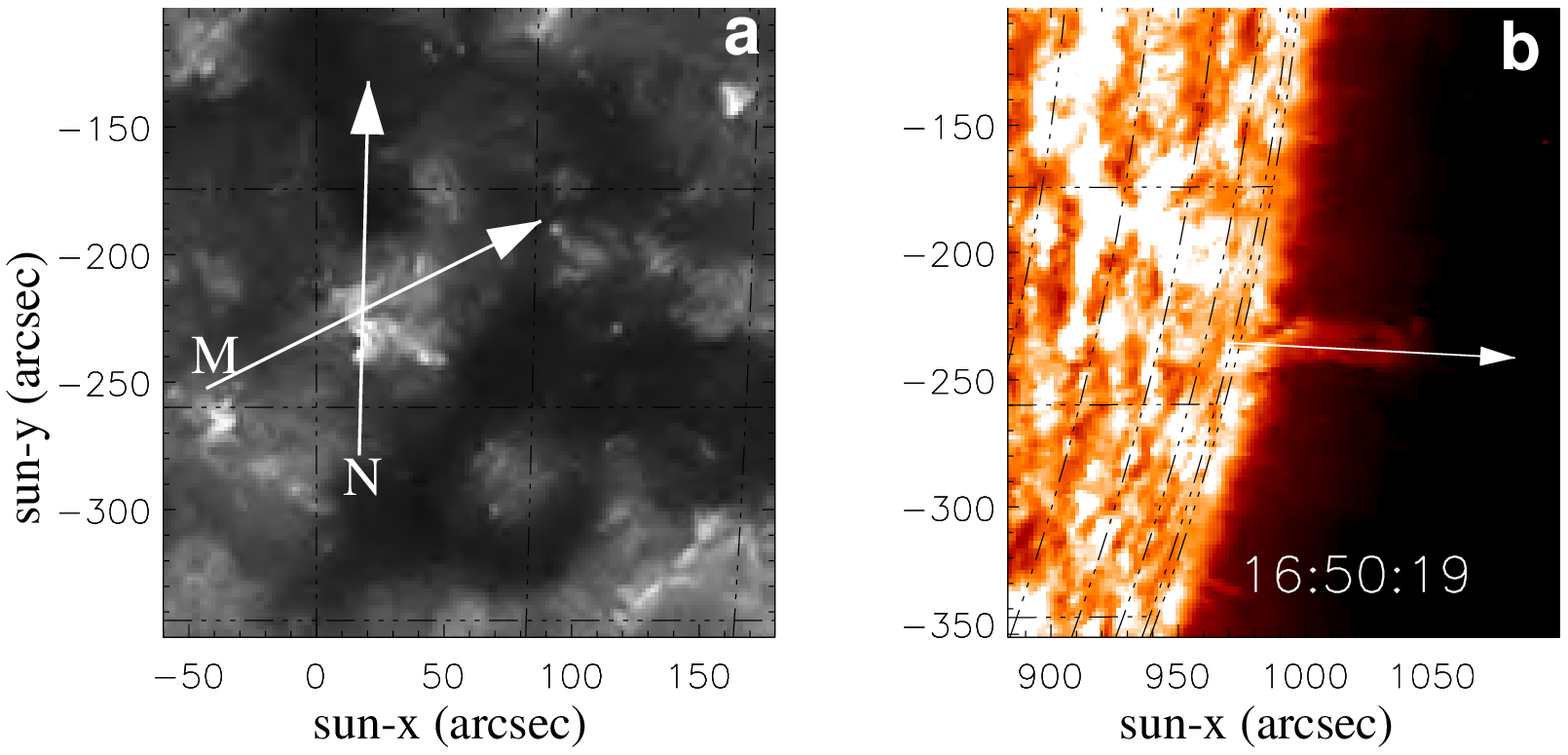}
\caption{The coronal hole eruption (a) 171\AA\ intensity (b) 304\AA\ intensity. The grid spacing is 5\degree. The arrows indicate the position and direction of time series in Fig.~\ref{14_ts_fig}. The evolution is shown in the on-line movies \href{http://www.mps.mpg.de/data/outgoing/innes/dims/vcu_ch.gif}{\tt ch.gif} and
\href{http://www.mps.mpg.de/data/outgoing/innes/dims/vcu_ch_r.gif}{\tt ch\_r.gif}} \label{14_vcu_fig}
\end{figure}

\begin{figure}
\centering
\includegraphics[width=9cm,height=3cm]{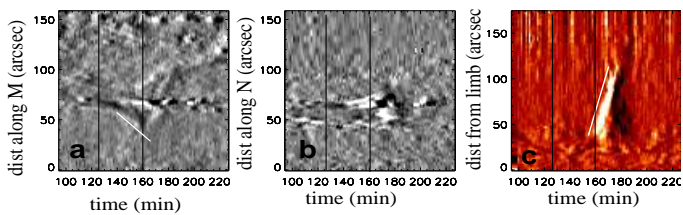}
\caption{Running difference time series along the arrows in Fig.\ref{14_vcu_fig}: (a) 171\AA\ along M (b) 171\AA\ along N (c) 304\AA. The black vertical lines are drawn at the same time in each image. They mark the beginning of the 171\AA\ dimming and the 304\AA\ eruption. White lines show fronts listed in Table 1.} \label{14_ts_fig}
\end{figure}

 The event discussed here and shown in Fig.~\ref{14_vcu_fig} and the movies 
  is the biggest seen in the coronal hole during 14 Feb observations. It occurred along a bright EUV ridge (Fig.~\ref{14_av_fig} and Fig.~\ref{14_vcu_fig}). Unlike the quiet Sun eruption, there was no strong EUV dimming. A weak dimming, expanding along the ridge from the eruption site, can be seen in the running difference movie. It is more obvious moving south-east, along arrow M, in the time series shown in Fig.~\ref{14_ts_fig}a. This image also reveals a faint front moving north-east along the ridge. We also notice that along the arrow N, pointing into the darkest parts of the coronal hole, there is no obvious front. Thus it seems as if the fronts move along the brighter channel running down the center of the coronal hole where there are presumably some mixed-polarity fields.

At the limb in 304\AA\ the eruption is jet-like and, compared with the quiet Sun event, an insignificant amount of material falls back to the disk. This is confirmed by the 304\AA\ running difference time series (Fig.~\ref{14_ts_fig}c) which shows no second brightening (c.f. Fig.~\ref{12_ts_fig}b).

\begin{table*}
\begin{threeparttable}
\caption{Start time and velocity of eruption fronts.}
\label{table1}
\centering
\begin{tabular}{l c c c c c c c c} \\
\hline\hline
 Fronts\tnote{1}&QS\_J\_171 & QS\_K\_171 & QS\_304 & PR\_Y\_304 & PR\_Z\_304 & CH\_M\_171 & CH\_N\_171 & CH\_304 \\
\hline
time (min) & 23& 32 & 32 & 15& 25 & 126 & 126 & 158\\
vel (\kms)& 28 & 9 & 55& -14& 12 & 12 & -- & 55 \\
    \hline
\end{tabular}
\begin{tablenotes}
\item[1]The names are made up of QS (Quiet Sun), {PR (Prominence)} or CH (Coronal Hole) followed by the arrow letter and then the wavelength.
\end{tablenotes}
\end{threeparttable}
\end{table*}

\section{Discussion}
We have investigated two chromospheric eruptions, seen simultaneously at the limb in 304\AA\ and at disk center in 171\AA.
 Measured onset times and velocities of dimming fronts and the corresponding chromospheric ejections are listed in Table 1. The velocities are between 10-30~\kms\ across the disk, similar to the values in \citet{Innes09}, and about 50-60~\kms\ into the corona. In both cases dimming in the coronal 171\AA\ emission precedes the chromospheric eruption indicating that coronal restructuring is important in the triggering mechanism of both quiet Sun and coronal hole eruptions.
We also found that about 5~min before the quiet Sun eruption, structures in a prominence about 100\arcsec\ along the limb, partially disrupted. The coronal hole dimming was essentially confined to the bright ridge running across the coronal hole.

\begin{figure}
\centering
\includegraphics[width=6.5cm]{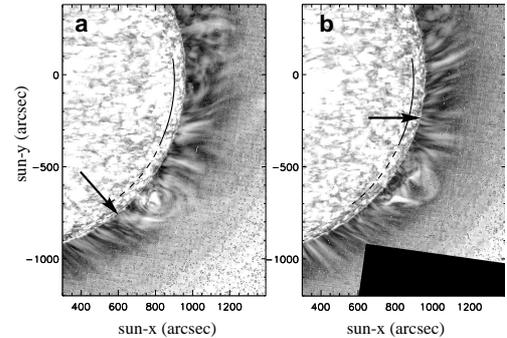}
\caption{Power maps of low frequency ($\le1$~mHz) 171\AA\ intensity for (a) 14:00-18:00 UT 12 Feb 2009 and
(b) 14:00-18:00 UT 14 Feb 2009. The position of the coronal hole (solid) and prominence (dashed) structures in Fig.~\ref{14_av_fig} are indicated. The arrows point to the positions of the chromospheric eruptions.} \label{power_plot}
\end{figure}

To relate these observations to the coronal field, we
 experimented with techniques to reveal its structure above the limb and found that maps representing the 171\AA\ intensity variation showed well organized structure with large closed arcades above prominences, disorganized patches above the quiet Sun and outwardly directed plumes above coronal holes.
In Fig.~\ref{power_plot} we show the low frequency 171\AA\ power maps for the 12 Feb 2009 and 14 Feb 2009 constructed using the co-temporal 171\AA\ STEREO-B images taken with a cadence of 150~s.
 The change in structure where the coronal hole reaches the limb on the 14~Feb is easily recognizable, and so is the arcade above the prominence. Arrows point to the positions of the two eruptions discussed. The quiet Sun eruption on the 12 Feb was on the outer edge of the enclosing arcade of the prominence (Fig.~\ref{power_plot}a). This could explain why a partial disruption of the prominence coronal field preceded coronal dimming and disruption in the connected quiet Sun.

The evidence presented here points to the triggering of these small-scale eruptions via the opening of the coronal magnetic field. {In this unique dataset, we see much interesting activity along the west limb in STEREO-B images; however these two events were special because they were big enough to be seen distinctly in 304\AA\ with no background confusion. Future high cadence, multi-filter observations with SDO/AIA of coronal dimmings associated with solar eruptions on all scales are expected to verify the frequency of
coronal re-structuring in the initiation of both large and small solar eruptions.}

\begin{acknowledgements}
The EUVI data were produced by an international consortium of NRL (USA), LMSAL (USA), GSFC (USA), RAL (UK), UB (UK), MPS (Germany), CSL (Belgium), IOTA (France), and IAS (France). STEREO is a project of NASA. This work was started at a workshop in ISSI, Bern "Small-scale transient phenomena and their contribution to coronal heating". NCAR is sponsored by the National Science Foundation.
\end{acknowledgements}

\bibliographystyle{aa}


\end{document}